\documentclass[12pt]{article}
\usepackage{graphicx}

\newcommand\pubnumber{NuPhys2017-Burns}
\newcommand\pubdate{\today}

\def\napoli{Atomic Weapons Establishment\\
Aldermaston, Reading, RG7 4PR, UNITED KINGDOM}
\def\support{\footnote{Work supported by the US, Department of Energy and the UK Ministry of Defence and Science and Technologies Facilities Council. 
\textcopyright British Crown Owned Copyright 2018/AWE}}

\def\Title#1{\begin{center} {\Large #1 } \end{center}}
\def\Author#1{\begin{center}{ \sc #1} \end{center}}
\def\Address#1{\begin{center}{ \it #1} \end{center}}

\newcommand\pubblock{\rightline{\begin{tabular}{l} \pubnumber\\
         \pubdate  \end{tabular}}}
\newenvironment{Abstract}{\begin{quotation}  }{\end{quotation}}
\newenvironment{Presented}{\begin{quotation} \begin{center} 
             PRESENTED AT\end{center}\bigskip 
      \begin{center}\begin{large}}{\end{large}\end{center} \end{quotation}}
\def\Acknowledgements{\bigskip  \bigskip \begin{center} \begin{large}
             \bf ACKNOWLEDGEMENTS \end{large}\end{center}}

\def\beq{\begin{equation}}
\def\eeq#1{\label{#1}\end{equation}}
\def\eeqn{\end{equation}}

\def\beqa{\begin{eqnarray}}
\def\eeqa#1{\label{#1}\end{eqnarray}}
\def\eeqan{\end{eqnarray}}

\let\bar=\overbar

\def\Dslash{\not{\hbox{\kern-4pt $D$}}}
\def\dslash{\not{\hbox{\kern-2pt $\del$}}}

\def\msb{{\bar{\ssstyle M \kern -1pt S}}}

\begin{document}
\begin{titlepage}
\pubblock

\vfill
\Title{Remote detection of undeclared nuclear reactors using the WATCHMAN detector}
\vfill
\Author{ Jonathan Burns for the WATCHMAN collaboration\support}
\Address{\napoli}
\vfill
\begin{Abstract}
Remote detection of undeclared nuclear reactors remains one of the key goals concerning global nuclear security. 
To meet this goal the WATCHMAN collaboration has proposed the construction of a water based antineutrino detector, 
sited 13 to 25 kilometres from a nuclear reactor complex. Antineutrinos from the reactor interact in the water of the 
detector via an inverse beta decay interaction resulting in two distinct cones of Cherenkov light tens of milliseconds apart. 
Using this interaction WATCHMAN (WATer Cherenkov Monitor for ANtineutrinos) will be the first detector to determine the active/inactive 
status of a reactor complex at a stand-off greater than 10 kilometres. The water used in the detector will be doped with gadolinium, 
providing the first demonstration of the potential of gadolinium doped detectors for reactor monitoring and will confirm the 
potential of the technology for use in larger multi-kiloton neutrino experiments. The proposed WATCHMAN design will be a kiloton scale 
water based detector, 
constructed of a 16 metre diameter tank with a height of 16 metres and will comprise approximately 3000 photomultiplier tubes.

An overview of the remote monitoring goals of the WATCHMAN collaboration will be given, with a detailed description 
of the proposed detector. An outline of the two proposed WATCHMAN sites will also be detailed with a prediction of the expected 
antineutrino rate and the time taken to determine the switch between the associated reactor on and off state at each site. 
A brief summary of the project and the future goals non-proliferation goals of the collaboration will also be presented.

\end{Abstract}
\vfill
\begin{Presented}
NuPhys2017, Prospects in Neutrino Physics
Barbican Centre, London, UK,  December 20--22, 2017
\end{Presented}\vfill
\end{titlepage}
\def\thefootnote{\fnsymbol{footnote}}
\setcounter{footnote}{0}

\section{Introduction}

The technical goal of the WATer CHerenkov Monitor of ANtineutrinos (WATCHMAN) demonstration is to detect antineutrinos from a 
reactor using a kiloton-scale water-based detector. WATCHMAN will be the first demonstration of remote monitoring of reactor operations 
at significant scale and standoff using scalable gadolinium doped water-based technology. The demonstration consists of observing 
reactor operations, including transitions from the reactor switch on and off (or vice-versa), using a 1000 ton gadolinium doped 
water target (fiducial mass) \cite{dazeley},\cite{bernstein}. 

\section{Boulby Proposed Site}

\begin{figure}[htb]
\centering
\includegraphics[width=5.5cm]{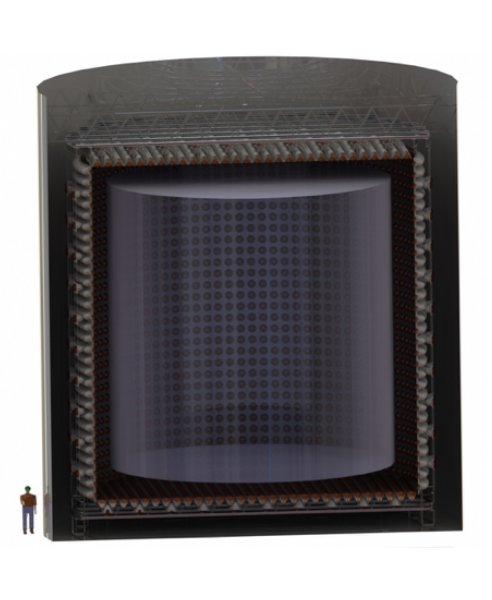}
\caption{a graphic illustrating the current design of the WATCHMAN detector}
\label{fig_detector}
\end{figure}

The WATCHMAN detector, see Figure \ref{fig_detector} will mark the first use of the chosen site's cavern and detector infrastructure for 
antineutrino detection. The facility can also accommodate different antineutrino target media, different 
photosensors and other changes to the baseline detector. This site and infrastructure is referred to as 
the Advanced Instrumentation Test-bed. The present summary serves as a record of the site selection for 
the Advanced Instrumentation Testbed and the WATCHMAN demonstration. 

In order to select a site for the demonstration, the collaboration conducted an exhaustive comparative 
analysis of possible deployment sites in the United States and the United Kingdom, taking into account 
trade-offs between detector size, standoff distance, cost, schedule and other factors. Within the cost 
and schedule envelope provided by our funding agencies, two criteria dominated our site selection review. 
The first was proximity of the deployment site to an existing reactor at ~5-50 km standoff distances, and 
the second was access to existing underground cavern space in order to greatly suppress backgrounds arising 
from cosmic rays, and to keep costs within the project budget. These two criteria alone reduced the number of viable 
candidate locations to two: the Boulby underground science facility, at 25 kilometres from the Hartlepool two-reactor 
complex in northern England, and the Morton salt mine, about 13 kilometres from the single reactor at the Perry site, 
near Lake Erie in the state of Ohio.  After an extensive comparative analysis of these two sites, we have selected the Boulby mine in the 
United Kingdom as the most appropriate location for the demonstration of this novel remote reactor monitoring technology. 

While the project goals are achievable at either site, Boulby has the following advantages:
\begin{itemize}
\item Greater standoff distances (25 km compared to 13 km);
\item A long-established and ongoing record of safe operation of an underground scientific facility, 
including extensive support infrastructure for scientific operations both below-ground, and the availability of 
above-ground support offices near the mine; 
\item A more complex reactor cycle, consisting of two reactors turning off and on with a variable outage schedule. 
This complexity allows for additional insight into the problems facing real-world monitoring situations, where it may be 
desirable to detector exclude the existence of an undeclared reactor in the presence of a declared reactor whose power, 
operational cycle and standoff are known.
\end{itemize}

The Morton Salt mine had its own advantages, including easier access for our collaborators based in the United States, a 
shorter dwell time needed to see the ~10x stronger signal that is available from the closer and more powerful reactor, 
and a shorter distance from the mine elevator to the deployment location. However, taking all relevant factors into account, 
the advantages of the Boulby site greatly outweighed those of the Morton Salt site.

\section{Reactor observation Time}

\begin{figure}[htb]
\centering
\includegraphics[width=8cm]{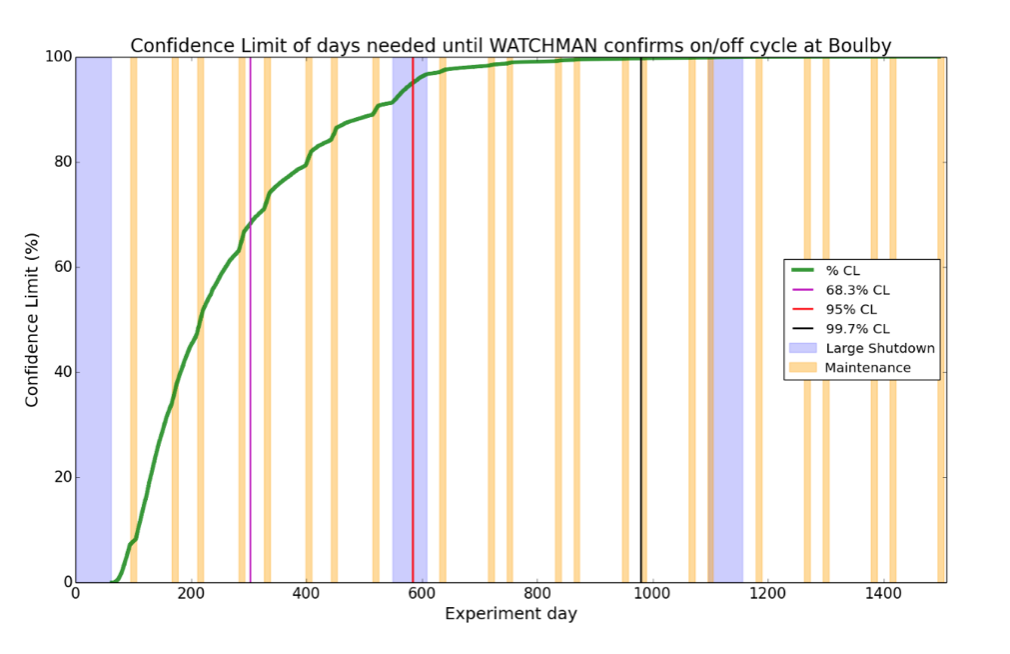}
\caption{The confidence level as a function of dwell time for observation of a difference between the one-reactor-on and two-reactor-on states, 
for the case of the Boulby site and the Hartlepool reactor complex.  Our observation goal is met once a three sigma difference between the 
one-reactor on and two-reactor-on data is observed, in 95\% of all experiments. As seen in the Figure, just under 600 days of data must 
be accumulated to meet this criterion (the vertical red line).}
\label{fig_reactor_times}
\end{figure}

A key criterion for success at either site is the ability to observe reactor operations within the time available for the project. 
We have found that both sites meet this criterion. With prior knowledge of both reactors outage schedules, we divide data into two categories: two reactors on, each at 100\% thermal power, 
and one reactor on at 100\% thermal power (rare two-reactor-off periods are neglected in our analysis). The one-reactor-on data 
contains events generated by antineutrinos arising from a single reactor - along with an irreducible background arising from 
distant reactor and non-antineutrino backgrounds that mimic the reactor signal. The two-reactor signal is defined similarly 
when both reactors are fully on. Based on our current efficiency and background estimates, just under 600 days are required 
at the Boulby site to observe a 3 sigma deviation between the two-reactor and one-reactor on states, for 95\% of all trial 
experiments. Note that this is a conservative standard: it implies that the average trial experiment predicts a deviation of 
4.6 sigma in 600 days. Figure \ref{fig_reactor_times} shows the evolution of our confidence level as we accumulate data over the course 
of the Hartlepool reactor cycles. The reactor outage schedule assumed here was modelled to reproduce the average outage times, 
including short maintenance outages and longer refuelling outages, reported by the Hartlepool facility over the last 3 years.

Based on this study, we could complete this 600 day run by about mid-2023, well before a nominal mid-2024 end of reactor 
operations at the Hartlepool site.  We are confident of success in achieving this sensitivity goal for a number of reasons.  
First, drawing on significant experience within the collaboration with water-based detectors and antineutrino detectors in 
general, we have invested a significant amount of effort in building a well-validated and tested GEANT4 \cite{geant1},\cite{geant2} simulation
for the purpose of accurately estimating our efficiencies and backgrounds. 
Second our simulation is not yet fully optimised; shorter dwell times may be achievable by adjusting the fiducial 
volume of the detector, by other analysis improvements, or by making other modest and low-cost changes to the detector design. 
Third,  the Hartlepool reactor complex is may continue operating beyond its 
nominal shutdown date, perhaps for an additional 3-8 years. This would provide significantly more time to achieve our baseline goal should 
there be any construction or other delays. 

Taken together, the 600 day dwell time value, the likely extension of reactor operations, and the possibility for further 
optimisation our analysis and of the detector design all show that the Boulby site is the optimum choice for achieving the project goals.  

\section{Conclusions}

In summary, the Boulby site offers an excellent opportunity to study the response of a large water-based detector to reactor 
operations at significant standoff.  Our initial and conservative criterion for observation of reactor operations can be met within 
our proposed timeline, and a more realistic criterion is being developed which is likely to shorten the time required to for 
successful observation of a transition in reactor operations. Since our other top-level criteria strongly favour the Boulby site, 
we have chosen this site for the Advanced Detector Instrumentation Testbed and the WATCHMAN demonstration.

\Acknowledgements
The author wishes to acknowledge the US and UK governments in supporting this work.


\begin{thebibliography}{99}

\bibitem{dazeley}
S. Dazeley, A. Bernstein, N. Bowden, and R. Svoboda, Observation of Neutrons with a Gadolinium Doped Water Cerenkov Detector, Nuclear Instruments and Methods A, 607, 616?619 (2009).

\bibitem{bernstein}
A. Bernstein, T. West, and V. Gupta, An assessment of antineutrino detection as a tool for monitoring nuclear explosions,Science \& Global 
Security 9, no. 3, 235-255 (2001).

\bibitem{geant1}
S. Agostinelli et al., Geant4 - A Simulation Toolkit,  Nuclear Instruments and Methods A 506, 250-303 (2003)

\bibitem{geant2}
J. Allison et al., Geant4 Developments and Applications,  IEEE Transactions on Nuclear Science 53 No. 1  270-278 (2006)






\end{thebibliography}
\end{document}